\documentclass[aps,prl,showpacs,twocolumn]{revtex4}

\usepackage{amsmath}
\usepackage{graphicx}
\usepackage{dcolumn}
\usepackage{bm}

\begin{document}

{\bf Reply to Valla's Comment on ``Multiple Bosonic Mode Coupling
in Electron Self-Energy of (La$_{2-x}$Sr$_x$)CuO$_4$"}\\

In his comment\cite{Valla}, Valla simulated the effect of energy
resolution on the fine structure in the electron self-energy. By
employing one energy resolution of $\Delta$$_{exp}$=20 meV in his
simulation, he concluded that the fine structure in Re$\Sigma$
cannot be observed and the fine structure we identified\cite{Zhou}
are most likely noise related.

First, let us look at data measured with an energy resolution
better than 20meV\cite{Challenge}.  Fig. 1 shows the measured real
part of electron self-energy Re$\Sigma$ taken at an energy
resolution of $\sim$15meV for LSCO x=0.03 (Fig. 1a)\cite{Combine}
and at 12 meV for LSCO x$\sim$0.06 (Fig. 1b). Clear fine
structure, manifested as a curvature change, can be identified at
40$\sim$45 meV and $\sim$60meV.  The measured Re$\Sigma$ is fitted
by the same procedure as in\cite{Zhou}. The second derivative of
the fitted Re$\Sigma$ reveals two additional weaker features at
$\sim$25meV and $\sim$80meV.

The new data with better energy resolution are consistent with
other data presented in \cite{Zhou}. The two features at
40$\sim$45 and $\sim$60meV are rather robust, showing up as main
features in all the measurements. The higher energy feature at
70$\sim$80 mev shows up more clearly in LSCO samples with higher
doping, same as before\cite{Zhou}. The variation of the low energy
feature at $\sim$25 meV is relatively larger among different
measurements, but the feature exists. Given consistent
observations from different samples, different dopings, and under
different experimental conditions, our findings are not
explainable by random noise. As shown in \cite{Zhou} as well as in
Fig. 1, experimental uncertainty including noise may affect the
results, such as the small peak shift between 40 and 45 meV.
However, bearing in mind the extreme challenging nature of this
experiment, we consider the consistency satisfactory in this very
first effort to get the fine structure in cuprates.

\begin{figure}[tbp*]
\begin{center}
\includegraphics[width=.85\columnwidth,angle=0]{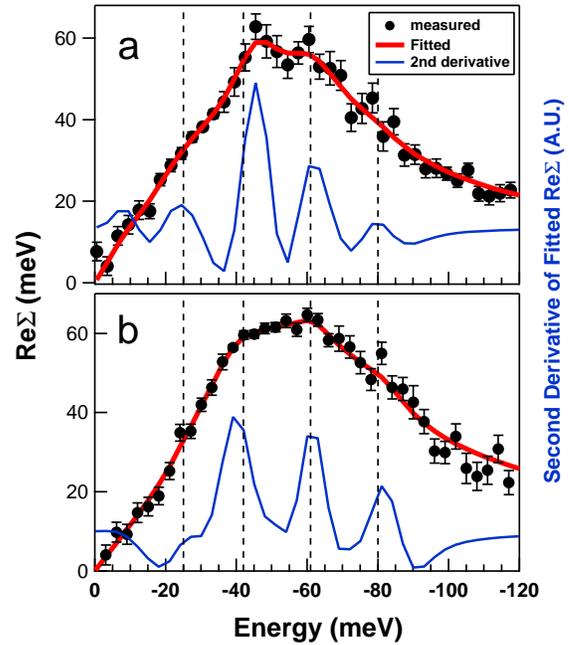}
\end{center}
\caption{Measured real part of the electron self-energy (solid
circles) for LSCO x=0.03 (a) and $\sim$0.06 (b) at 20K. The total
energy resolution for LSCO x=0.03 (a) is around 15 meV and for
LSCO x=0.06 (b) is around 12 meV. The red solid lines represent a
fitting of the measured Re$\Sigma$. The second derivative of the
fitted data is shown as blue solid lines. }
\end{figure}

We now turn to the issue raised by Valla\cite{Valla} about the
effect of energy resolution on the fine structure.  First we note
that this simulation is model dependent because a specific
electron-phonon coupling form was used\cite{Valla}.  We also note
that this simulation is parameter sensitive, including the mode
energy, mode strength, impurity scattering and how the resolution
is included.  We don't think that one should take the simulation
too literally as in [1].  For the sake of discussion, let us put
these aside for the moment.  We show in Fig.2  simulated
Re$\Sigma$ under different energy resolutions in the same
procedure as in \cite{Valla}(Fig. 2a),  and the corresponding
second derivative (Fig. 2b). Four modes at 25, 40, 60, and 80meV
are assumed in the $\alpha$$^{2}$F($\omega$) to calculate the
single particle spectral function. The mode positions are
determined from Fig. 1b which were measured using a better energy
resolution, and an impurity term $\sim$100meV from MDC width at
E$_F$ is added in the imaginary part of the self-energy.

\begin{figure}[floatfix]
\begin{center}
\includegraphics[width=1\columnwidth,angle=0]{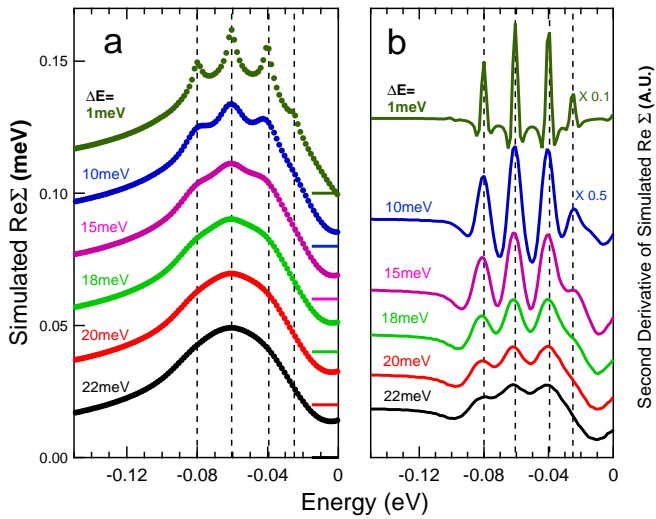}
\end{center}
\caption{(a). Simulated real part of electron self-energy
Re$\Sigma$ at different energy resolutions.  (b). Second
derivative of the simulated Re$\Sigma$ at different energy
resolutions.
 }
\end{figure}

As seen from Fig. 2, at a high energy resolution, the fine
structure shows up as small peaks on the top of a broad
self-energy feature. As the resolution deteriorates, the fine
structure manifests themselves as a curvature change in Re$\Sigma$
which can be better identified as peaks in its corresponding
second derivative. It is clear from Fig. 2b that at an energy
resolution of 20 meV or even slightly worse, it is possible to
identify the fine structure, albeit with a larger error bar. We
stress that while it appears hard to discern features in the
Re$\Sigma$ at an energy resolution of 20 meV, as claimed by
Valla\cite{Valla},  the second derivative that we used, reveals
them. The main reason here is that the fine structure is embedded
in the curvature of the self-energy. Part of the issue is the
separation between 45 and 61 meV mode, which is under an energy
resolution of 20 meV\cite{Valla}. The reason they are observable
can be three-fold.  As in Fig.2, one can still discern peaks in
the second derivative even when the separation is slightly smaller
than the resolution.  The second is that the energy resolution
quoted in our paper (18$\sim$20meV)\cite{Zhou} was an overestimate
as the beamline is conservatively calibrated. The third is a few
meV shift of the 40 meV mode to 45 meV as a small experimental
uncertainty is not unexpected.

A separate issue Valla raised is the distortion of Re$\Sigma$ at
lower energy near the Fermi level E$_F$\cite{Valla}. At a low
energy resolution,  the simulated Re$\Sigma$ exhibits a backbend
near E$_F$ and gives rise to a finite value at E$_F$. For example,
at an energy resolution of 20meV, the Re$\Sigma$ tends to level
off in the energy range between 0 and $\sim$20meV.  This simulated
result deviates markedly from our experimental results where the
measured Re$\Sigma$  extends to Fermi level almost linearly(Fig.
1).  This discrepancy between data and simulation is well beyond
the noise.  This issue depends on many details of the simulation
and its comparison with data that are beyond this short
communication.  However, the clear discrepancy between data and
simulation drives home an important point.  The real way for the
author of \cite{Valla} to substantiate his claim is not through a
model specific, parameter and procedure sensitive simulation, but
rather through experimental data with superior energy resolution
and signal to noise ratio.

In summary, experiments with improved energy resolution more
clearly reveals the fine structure of mode coupling which is
consistent with other measurements. Even within the context of
Valla's simulation, at an energy resolution of 20meV, it is
possible to observe fine structure as they are embedded in the
curvature and will be revealed in the second derivative. We
believe that the ensemble of data and the analysis make a strong
case for multiple mode coupling in LSCO, with the most important
modes near 40 and 60 meV.
\\
\\
\\
X. J. Zhou$^{1,2}$, Junren Shi$^{3}$, W. L. Yang$^{1,2}$, Seiki Komiya$^{4}$, Yoichi Ando$^{4}$, W. Plummer$^{3}$, Z. Hussain$^{2}$, and Z.-X. Shen$^{1}$\\
\\
\\
$^{1}$Dept. of Physics, Applied Physics and SSRL, Stanford University,  Stanford, CA 94305\\
$^{2}$Advanced Light Source, LBNL, Berkeley, CA94720\\
$^{3}$Condensed Matter Sciences Division, Oak Ridge National
Laboratory, Oak Ridge, TN37831.\\
$^{4}$CRIEP, Tokyo, Japan\\



\begin{thebibliography}{99}
\bibitem{Challenge} This experiment is challenging because of the necessity to use a relatively high
photon energy ($\sim$55eV) for LSCO system. In this case, it is
hard to compromise between two conflicting requirements: high
energy resolution and high photon flux for a good statistics of
data.
\bibitem{Combine} This is the same data set as in Fig. 2(a2) in
\cite{Zhou}. The data were measured at an energy resolution of
$\sim$15meV with 1 meV energy interval. To improve the statistics,
every three adjacent data point are combined to obtain one data
point in Fig. 1a.
\end{thebibliography}
\end{document}